\documentclass[useAMS,usenatbib]{mn2e}
\usepackage{subfigure}
\usepackage{rotating}
\usepackage{times}
\usepackage{graphicx}
\usepackage{lscape}

\title[GRB 051103]{A new analysis of the short-duration, hard-spectrum GRB 051103, a possible extragalactic SGR giant flare}

\author[Hurley et al.]
{K. Hurley$^{1}$\thanks{E-mail:khurley@ssl.berkeley.edu}, A. Rowlinson$^{2}$, E. Bellm$^{1}$, D. Perley$^{3}$, I. G. Mitrofanov$^{4}$, 
D. V. Golovin$^{4}$,
\newauthor
A. S. Kozyrev$^{4}$, M. L. Litvak$^{4}$, A. B. Sanin$^{4}$, W. Boynton$^{5}$, 
C. Fellows$^{5}$, K. Harshmann$^{5}$,
\newauthor
M. Ohno$^{6}$, K. Yamaoka$^{7}$, Y. E. Nakagawa$^8$, D. M. Smith$^{9}$, T. Cline$^{10}$, N.R. Tanvir$^{2}$,
\newauthor
 P.T. O'Brien$^{2}$, K. Wiersema$^{2}$, E. Rol$^{2,11}$, A. Levan$^{12}$, J. Rhoads$^{13}$, A. Fruchter$^{14}$,
\newauthor
 D. Bersier$^{15}$, J.J. Kavelaars$^{16,17}$, N. Gehrels$^{10}$, H. Krimm$^{10}$, D. M. Palmer$^{18}$, R. C. Duncan$^{19}$, 
\newauthor
C. Wigger$^{20}$, W. Hajdas$^{20}$, J.-L. Atteia$^{21}$, G. Ricker$^{22}$, R. Vanderspek$^{22}$, A. Rau$^{23}$, and
\newauthor
A. von Kienlin$^{23}$
\\
$^1$University of California, Space Sciences Laboratory, 7 Gauss Way, Berkeley, CA 94720-7450, U.S.A. \\
$^{2}$Department of Physics \& Astronomy, University of Leicester, University Road, Leicester, LE1 7RH, UK\\
$^3$Department of Astronomy, University of California, Berkeley, CA 94720-3411, U.S.A. \\
$^4$Institute for Space Research, Profsojuznaja 84/32, Moscow 117997, Russian Federation \\
$^5$University of Arizona, Lunar and Planetary Laboratory, Tucson, AZ 85721, U.S.A.\\
$^6$Institute of Space and Astronautical Science (ISAS/JAXA), 3-1-1 Yoshinodai, Sagamihara, Kanagawa 229-8510, Japan \\
$^7$Department of Physics and Mathematics, Aoyama Gakuin University, 5-10-1 Fuchinobe, Sagamihara, Kanagawa, 229-8558, Japan \\
$^8$The Institute of Physical and Chemical Research (RIKEN), 2-1 Hirosawa, Wako, Saitama 351-0198, Japan \\
$^9$Department of Physics and Santa Cruz Institute for Particle Physics, University
of California, Santa Cruz, 1156 High Street , Santa Cruz, CA 95064, U.S.A. \\
$^{10}$NASA Goddard Space Flight Center, Code 661, Greenbelt, MD 20771, U.S.A. \\
$^{11}$ University of Amsterdam, Science Park Amsterdam, Kruislaan 403, 1098 SJ, Amsterdam \\ 
$^{12}$ Department of Physics, University of Warwick, Coventry CV4 7AL\\ 
$^{13}$ School of Earth and Space Exploration, Arizona State University, Tempe, AZ 85287, USA\\ 
$^{14}$ Space Telescope Science Institute, 3700 San Martin Drive, Baltimore, MD 21218, USA\\ 
$^{15}$ Astrophysics Research Institute, Liverpool John Moores University, Twelve Quays House, Birkenhead, CH41 1LD, UK\\ 
$^{16}$ Herzberg Institute of Astrophysics, National Research Council, 5017 West Saanich Road, Victoria, BC V9E 2E7 \\ 
$^{17}$ Visiting Astronomer, Kitt Peak National Observatory, National Optical Astronomy Observatory, which is operated by the Association of Universities \\
$^{18}$Los Alamos National Laboratory, P.O. Box 1663, Los Alamos, NM 87545, U.S.A. \\
$^{19}$The University of Texas at Austin, Department of Physics, 1 University Station C1600, Austin, Texas  78712-0264, U.S.A.\\
$^{20}$Paul Scherrer Institute, 5232 Villigen PSI, Switzerland\\
$^{21}$Laboratoire d'Astrophysique, Observatoire Midi-Pyr\'{e}r\'{e}es,
14 avenue E. Belin, 31400 Toulouse, France\\
$^{22}$Kavli Institute for Astrophysics and Space Research, Massachusetts Institute of Technology,
70 Vassar Street, Cambridge, MA 02139, U.S.A.\\
$^{23}$Max-Planck-Institut f\"{u}r extraterrestrische Physik,
Giessenbachstrasse, Garching, 85748 Germany
}

\begin{document}

\date{}

\pagerange{\pageref{firstpage}--\pageref{lastpage}} \pubyear{2002}

\maketitle

\label{firstpage}

\begin{abstract}
GRB 051103 is considered to be a candidate soft gamma repeater (SGR) extragalactic giant magnetar flare by
virtue of its proximity on the sky to M81/M82, as well as its time history, localization, and energy spectrum.  We have 
derived a refined interplanetary network 
localization for this burst which reduces the size of the error box by over a factor of two.  We examine its time
history for evidence of a periodic component, which would be one signature of an SGR giant flare, and conclude
that this component is neither detected nor detectable under reasonable assumptions.  We analyze the
time-resolved energy spectra of this event with improved time- and energy resolution, and conclude that although
the spectrum is very hard, its temporal evolution at late times cannot be determined, which further complicates
the giant flare association.
We also present new optical observations reaching limiting magnitudes of $R > 24.5$, about 4 magnitudes deeper than previously 
reported. In tandem with serendipitous observations of M81 taken immediately before and one month after
the burst, these place strong constraints on any rapidly variable sources in the region of the
refined error ellipse proximate to M81. We do not find any convincing afterglow candidates
from either background galaxies or sources in M81, although within the refined error region we 
do locate two UV bright star forming regions which may host SGRs. A supernova remnant (SNR) 
within the error ellipse could provide further support for an SGR giant flare association, but 
we were unable to identify any SNR within the error ellipse. These data still do not allow 
strong constraints on the nature of the GRB 051103 progenitor, and suggest that candidate 
extragalactic SGR giant flares will be difficult, although not impossible, to confirm.



\end{abstract}


\section{Introduction}

Giant flares are the most spectacular manifestations of soft gamma repeaters (SGRs).
Their time histories are characterized by a very rapid ($<$ 1 ms) rise to an intense
peak lasting several hundred milliseconds, followed by a weaker, oscillatory phase which exhibits the period
of the neutron star.  The spectrum of the peak is very hard and extends to MeV energies.
The most energetic giant flare to date is that of 2004 December 27 from SGR1806-20,
with an isotropic energy of well over 10$^{46}$ erg.
(Hurley et al. 2005, Palmer et al. 2005, Mereghetti et al. 2005, Terasawa et al. 2005,
Frederiks et al. 2007b).
The observation of this event raised the question once more
of the existence of extragalactic giant magnetar flares, and
their relation to the short cosmic gamma-ray bursts (GRBs).  Viewed from a large distance,
only the initial peak of a giant flare would be detectable, and it would resemble
a several hundred millisecond long, hard spectrum GRB.  The energetics
of giant flares make it a virtual certainty that such events can be
detected in distant galaxies, but recognizing them and demonstrating
their origin beyond a reasonable doubt remain difficult tasks.  Predictions
of the rates of extragalactic giant flares, and the percentage of
short cosmic bursts which might actually be giant flares, vary widely (from a few
to $\sim$ 15\% - Lazzati et al. 2005, Nakar et al. 2006, Popov and Stern 2006,
Tanvir et al. 2005), in part because of their unknown
number-intensity relation (no SGR has yet been observed to emit more than
one giant flare, and their distances are uncertain in most cases).  However, they generally agree that the rate
is small enough that the majority of short bursts are indeed not due to
SGR giant flares.  On the other hand, the rate is not expected to be zero, so it is important
to examine all possible cases exhaustively.  In this paper, we revisit GRB 051103, a short burst whose location,
time history, and energy spectrum are consistent with an origin as a giant
flare in M81.  We make use of the full interplanetary network (IPN) data set to
obtain a refined localization (an error ellipse).  We utilize the capability of the RHESSI
spacecraft to obtain time-resolved energy spectra with good energy resolution, 
at a time resolution which is limited only by counting statistics, and we take
advantage of the fact that three instruments recorded energy spectra with good statistics
to obtain joint fits.  Our analysis follows
that of Frederiks et al. (2007a), which was based on the Konus-Wind data.

We also present new, much deeper optical data than previously reported for the section of the refined error ellipse 
closest to M81, taken 3 days after GRB 051103 \citep[and approximately 16 hours after the ]
[GCN notice] {golenetskii2005}. We use these data to search for possible optical counterparts 
of this short burst (SGRB), and discuss the implications of its non-detection for its progenitor and 
putative association with M81.  Throughout this paper, we utilize the distance modulus
of M81 determined by HST observations of Cepheids, 
27.8 \citep[3.6 Mpc, ][]{freedman1994}.

\section{IPN observations and localization}

GRB 051103 occurred at 09:25:42 UT at Earth, and was observed by HETE-FREGATE (Atteia et al. 2003), 
RHESSI (The Ramaty High Energy Solar Spectroscopic Imager - Smith et al. 2002), Suzaku-WAM (Yamaoka et al. 2009), and Swift-BAT (Gehrels et al. 2004) in low-Earth orbit; the burst was outside the coded
fields of view of Swift-BAT and the HETE-II imaging instruments, and was therefore not localized by them.  
It was also
observed by INTEGRAL SPI-ACS (Rau et al. 2005) at 0.5 light-seconds from Earth, Konus-Wind (Aptekar et al. 1995) at
4.5 light-seconds from Earth, and by Mars Odyssey (HEND and GRS experiments, Hurley et al. 2006)
at 232 light-seconds from Earth.  A preliminary IPN error box was announced in a GCN Circular,
where it was pointed out that this event had the largest peak flux ever observed by
Konus-Wind for a short burst (Golenetskii et
al. 2005).
Optical follow-up observations were reported by Lipunov et al. (2005a,b), Klose
et al. (2005), and Ofek et al. (2005, 2006), and a radio observation was reported
by Cameron and Frail (2005).  All yielded negative results.
A MILAGRO GeV/TeV observation similarly produced only upper limits (Parkinson et al. 2005).

The observations of each statistically independent spacecraft pair can be analyzed to produce an annulus
of location, and the annuli can be combined to yield an error ellipse using
the method described in Hurley et al. (2000).  In this case, we have combined
the Konus-HETE, Konus-RHESSI, Konus-INTEGRAL, Konus-Swift, and Konus-Odyssey
annuli.  Under these conditions (several relatively short baselines and one
relatively long one), the error ellipse has a long major axis, 
corresponding to the annuli derived from the closer spacecraft pairs, and a
short minor axis, corresponding to the annulus from the distant spacecraft
pair.  We obtain a 3 $\sigma$ error
ellipse centered at $\alpha(2000)$=147.933$\degr$, $\delta(2000)$=+68.589$\degr$,
with major and minor axes 137\arcmin and 0.96\arcmin respectively, and area
104 square arcminutes.  The chi-square for the error ellipse center is 0.9
for 3 degrees of freedom (5 annuli minus 2 fitted coordinates).  The area of the initial error box was 240
square arcminutes\footnote{A typographical error in GCN 4197 incorrectly gave the
area as 120 square arcminutes}.  The initial error box and the final error ellipse are shown
in figure 1.

\begin{figure}
\centering
\includegraphics[width=8.2cm]{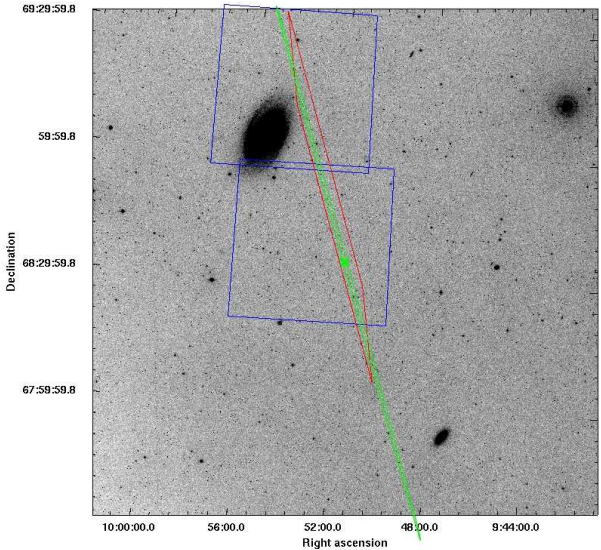}
\caption{This shows in red the original error trapezium provided by the IPN, in green 
the 3$\sigma$ refined error ellipse for the position of GRB 051103 and in blue the fields 
of the region studied using KPNO.
The asterisk indicates the center of the ellipse and the most likely arrival
direction of the burst.  Approximately 65 square arcminutes of the ellipse are
contained within the old error box. These are imposed upon an image of the area surrounding 
M81 from the Digital Sky Survey.  The 
possibility that this burst came from the inner disk of M81 is firmly ruled out.  However,
the brightest GALEX UV knots noted by Ofek et al. (2006) are within the
ellipse.  Lipunov (2005b) noted the presence of two galaxies within the initial error box,
PGC 2719634 and PGC 028505.  The former galaxy lies at the 18\% confidence contour of the
ellipse, and remains a plausible host candidate, while the latter lies at the 0.03\% contour,
and is unlikely to be the host.}
\end{figure}

\section{Time history}

The RHESSI time history of GRB 051103 is shown in the top panel of figure 2.
A distinctive signature of all three previously observed giant SGR flares within our Galaxy and the LMC to date is the periodic extended component following the initial short-duration peak.  Among these three events, the periods of this extended tail have clustered around a narrow range of 5--8 seconds and also have a relatively narrow range of total isotropic energy releases of $1-4 \times 10^{44}$ erg.  This signal lasts for many minutes following the bursts but falls off rapidly after a few hundred seconds.  While extended emission is frequently detected following cosmological short-hard bursts, such emission is not periodic.  Therefore detection of a periodic component of emission would be considered a strong confirmation of an SGR origin.

None of the IPN light curves shows obvious evidence for extended emission (pulsed or otherwise) following the burst.  However, it is conceivable that a marginally detected signal could be present within the noise.  To search for such a component, we acquired Swift-BAT data for GRB 051103 (binned at 64 ms) and used the Lomb (1976) periodogram to calculate the relative power in the signal following the burst at periods up to about 20 seconds.  We created periodograms for all of the four BAT energy channels, which cover the energy range 15 to 350 keV (and for combinations of channel sums) and for various time ranges following the emission (ranging from the first 60~s to the first 300~s.)  To assess the significance of any peaks in the power spectrum, we performed a Monte Carlo analysis by repeatedly randomizing the order of the 64 ms time bins for each data set over the range of interest and measuring the rate of occurrence of independent peaks above various power levels.  We identified no peaks with greater than 98\% significance in any channel or time range.

This non-detection is expected.  To assess the general detectability of periodic post-flare emission from extragalactic giant magnetar flares, we also acquired the Swift-BAT light curve of the the 2004 December 27th flare from SGR 1806-20.  We then scaled the signal down by a factor of $(D/D_{\rm SGR})^2$ and added it to the light curve of GRB 051103 (both scrambled and unscrambled).  No signal is detected in the periodogram at the known periodicity of 7.56 seconds at the distance of M81/82 ($D = 3600$ kpc).  The maximum distance for detecting periodicity with our analysis greater than 3 sigma is only about $D = 220$ kpc (if a distance of $D_{\rm SGR}=14.5$ kpc to SGR 1806-20 is assumed, or to 130 kpc if the 8.7 kpc distance of Bibby et al. 2008 is assumed), less than the distance even to M31.  This limit may not be exact:  both SGR1806-20 and GRB 051103 were detected off-axis by BAT and the comparative satellite sensitivity will depend on the specifics of the off-axis angle. For the giant flare from 1806-20, BAT was pointing 105\degr away, and slewed to 61\degr away starting around 38 s after the peak.  For 051103, BAT was pointing 122\degr from the source.  However, as the expected signal from a December 27-like event at the distance of M81/M82 would be only 0.01 sigma assuming similar sensitivities for the two events, we consider it extremely unlikely that any possible angle outside the BAT FOV would lead to a detection unless the periodic component were several orders of magnitude stronger than that observed in the three Galactic/LMC events to date.

\section{Energy spectrum}

A key signature of the spectra of the three SGR giant flares observed to date is a very hard energy spectrum for the initial, several hundred millisecond long burst, 
and a dramatic spectral evolution to a soft spectrum for the subsequent pulsating component.  As these bursts were observed by various instruments, with different temporal
resolutions, spectral resolutions, and energy ranges, and all of them were in some degree of saturation at the peak, a precise description of the spectra is impossible.
Nevertheless, all of them can be characterized as very hard spectra at the peak, sometimes consistent with a very high temperature blackbody (e.g. Mazets et al. 1979, Fenimore et al. 1981, Hurley et al. 1999, 
Mazets et al. 1999, Hurley et al. 2005, Frederiks et al. 2007b).  Accordingly, we have analyzed the time-resolved energy spectra of GRB 051103.
RHESSI, Konus, and Suzaku obtained energy spectra for GRB 051103 over a wide energy range, with good statistics, although with different time resolutions.  
(Due to the off-axis arrival angles at Swift and HETE-II, the detector response matrices
are not well known, and we have not used these data.)
Because the finest time resolution can be obtained from the RHESSI data, we have analyzed the RHESSI spectra
both separately, to obtain the best time resolution, limited only by counting statistics, and combined with the Konus and Suzaku data, to obtain
the best statistics, albeit at the cost of temporal resolution.

RHESSI uses 
nine unshielded coaxial germanium detectors to observe a broad energy band (30~keV--17~MeV) 
with excellent energy (1--5~keV) and time resolution (1 binary $\mu$s) and 
moderate effective area ($\sim$150~cm$^2$).  The data are recorded 
event-by-event, which provides great flexibility in choosing analysis intervals.

To determine RHESSI's spectral response to GRB 051103, we used the Monte
Carlo package MGEANT (Sturner et al. 2000).  We 
simulated monoenergetic photons in 192 logarithmic energy bins ranging from
30~keV--30~MeV generated along a 60\degr azimuthal arc at the 97\degr off-axis angle
of GRB 051103.  
We fit a polynomial background and extracted the burst data in SolarSoftWare-IDL routine\footnote{http://www.lmsal.com/solarsoft/}.
Because of radiation damage to some of the detectors, we used only data from
rear segments 1, 4, 6, 7, and 8.
Spectral fitting was conducted with ISIS v1.4.9 (Houck 2000).  In
general, the full 30~keV--17~MeV energy band was employed, except when
sufficient counts could not be accumulated at high energies.

We fit the data with a Band (Band et al. 1993) function:
\[
N_E = \left\{
\begin{array}{lr}
  A (E/E_{piv})^{\alpha} \exp(-E/E_0) & E<E_{break} \\
  B (E/E_{piv})^{\beta}               & E>E_{break}
\end{array}
\right.
\]
with $E_{break} \equiv E_{0} (\alpha - \beta)$ and
$B \equiv  A (\frac{(\alpha-\beta) E_0}{E_{piv}})^{\alpha-\beta}
\exp(\beta-\alpha)$.  For $\beta < -2$ and
$\alpha > -2$, $E_{peak} \equiv E_0 (2 + \alpha)$ corresponds to the peak of the
$\nu F_\nu$ spectrum.  The normalization $A$ has units photons/(cm$^2$ s
keV), and $E_{piv}$ is here taken to be 100 keV.
For joint fits, the Band function parameters $\alpha$,
$\beta$, and E$_{peak}$ were tied for both instruments,
but the normalizations were allowed to vary independently.

For the RHESSI-only time-resolved fits, we identified time intervals with
background-subtracted S/N of 20 in the 60~keV--3~MeV band.  This yielded
three intervals, to which a fourth tail interval of S/N $=$ 12 was added.
For most intervals, the cutoff power law model (CPL), equivalent to the Band
function below $E_{break}$, provided the best fit.
The time evolution of the parameters of the best-fit spectral model (a
cutoff power-law) is presented in the lower panels of Figure 2.  The 
initial spike of emission has a significantly higher peak energy than the
decaying tail; however, the spectral index of the power law appears to harden
throughout the burst.  The results are reported in table 1.

\begin{figure}
\includegraphics[scale=0.4]{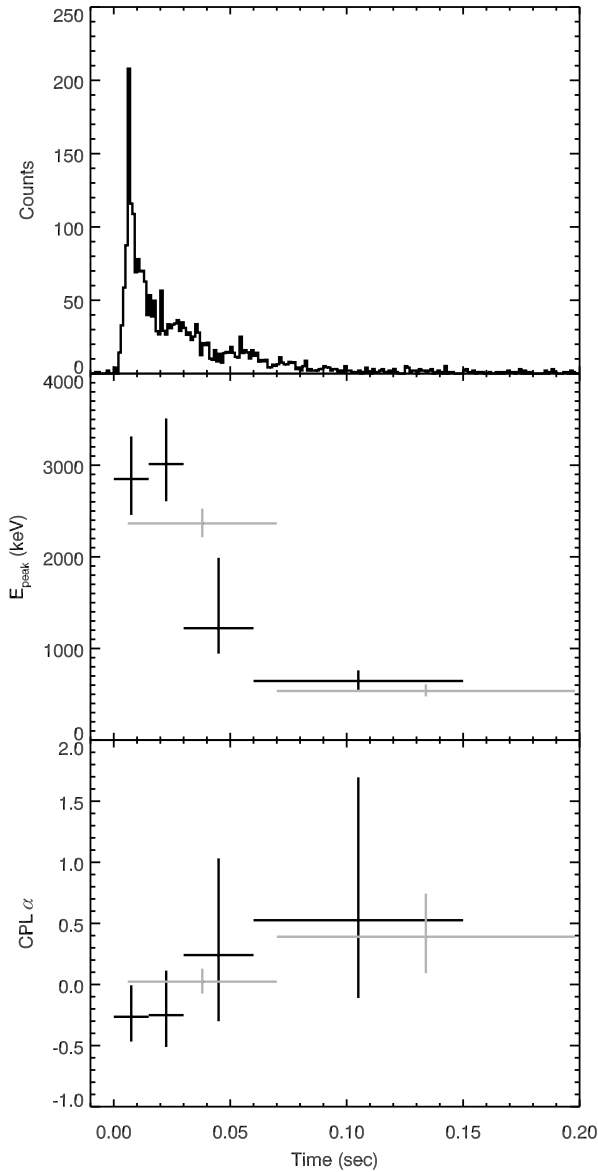}
\caption{Time history of GRB 051103, and evolution of the spectrum.  The top plot shows
the dead-time corrected RHESSI light curve (60~keV--3~MeV) with 1~millisecond time resolution, starting
at 09:25:42.184 UT.  The background count rate is 0.55 counts/ms and has not been
subtracted.  The time history has an e-folding rise time of 1.2 $\pm$ 0.04 ms, an e-folding
decay time of 28.6 $\pm$ 0.6 ms, and a T$_{90}$ of 100 $\pm$ 4 ms.
The middle and bottom plots show the evolution of the best-fit
peak spectral energy and power-law index for the cutoff power-law model.  
The black points are RHESSI-only, while the gray points are joint fits
between RHESSI and Konus-Wind.  }
\label{fig-evol}
\end{figure}

Suzaku-WAM did not trigger on GRB 051103, so the only data available are for a
1-second spectrum containing the entire burst, in the 50 keV -- 5 MeV energy range.  The RHESSI-WAM joint fit is shown
in Figure 3 and the fit results are reported in table 1.  The
WAM fluence is a factor of $\sim$5 lower than the RHESSI fluence; this
deficit appears to be a result of data lost due to deadtime during the intense peak 
of emission.

Konus-Wind triggered on GRB 051103 and recorded 64 ms spectra in the
20 keV -- 10 MeV range; we conducted joint fits between RHESSI
and Konus-Wind for the 64 ms and 128 ms intervals analyzed in Frederiks et
al. 2007a.   These fits are presented in Figures 3 and 4, and the details are
reported in table 1.  Good correspondence was obtained in the best-fit parameters
between the two instruments, although a normalization offset was necessary.

\begin{table*}
\begin{minipage}{140mm}
\caption{
Best fit parameters for the joint fits of the RHESSI/Suzaku-WAM, 
RHESSI/Konus-Wind, and RHESSI-only data. 
Times are relative to $T0 =$ 09:25:42.190 UT in the RHESSI frame.
Joint fit fluences are in the 20~keV--10~MeV band, while the RHESSI-only
fluences are 30~keV--10~MeV.  The instrument normalizations
were free to float in the fit; the normalization of the second instrument
relative to RHESSI is given. 
Errors are quoted at the 90\% confidence level.  
}
\begin{tabular}{@{}lccccccc@{}}
\hline
 Instruments & Interval &  $E_{peak}$ & $\alpha$ & $\beta$ & RHESSI Fluence         &  Normalization & $\chi^2$/dof \\
	     & (sec)    & (keV)       &          &         & (10$^{-5}$ erg/cm$^2$) &  Offset        &              \\
\hline

RHESSI + WAM & 
-0.011 -- 0.989 &
2235$^{+290}_{-280}$ &
-0.63$^{+0.11}_{-0.09}$ &
-2.59$^{+0.07}_{-0.41}$ &
4.80$^{+0.23}_{-0.23}$ &
0.196$^{+0.013}_{-0.012}$ &
44.7/37 = 1.21 \\

RHESSI + KW & 
0.000 -- 0.064 &
2080$^{+180}_{-200}$ &
0.13$^{+0.14}_{-0.11}$ &
-2.78$^{+0.31}_{-0.45}$ &
3.17$^{+0.18}_{-0.18}$ &
1.00$^{+0.09}_{-0.08}$ &
93.0/65 = 1.43 \\

RHESSI + KW & 
0.064 -- 0.192 &
536$^{+71}_{-59}$ &
0.39$^{+0.35}_{-0.30}$ &
--- &
0.156$^{+0.025}_{-0.024}$ &
1.23$^{+0.26}_{-0.20}$ &
30.9/32 = 0.96 \\

RHESSI & 
-0.006 -- 0.009 &
2850$^{+465}_{-390}$ &
-0.26$^{+0.26}_{-0.20}$ &
--- &
1.66$^{+0.18}_{-0.18}$ &
--- &
38.5/11 = 3.50 \\

RHESSI & 
0.009 -- 0.024 &
3010$^{+495}_{-405}$ &
-0.25$^{+0.36}_{-0.26}$ &
--- &
1.10$^{+0.12}_{-0.12}$ &
--- &
7.5/11 = 0.68 \\

RHESSI & 
0.024 -- 0.054 &
1220$^{+770}_{-280}$ &
0.24$^{+0.79}_{-0.54}$ &
--- &
0.66$^{+1.79}_{-0.18}$ &
--- &
3.7/4 = 0.93 \\

RHESSI & 
0.054 -- 0.144 &
645$^{+115}_{-95}$ &
0.53$^{+1.17}_{-0.64}$ &
--- &
0.145$^{+0.023}_{-0.022}$ &
--- &
17.8/7 = 2.54 \\

\hline
\end{tabular}
\end{minipage}
\end{table*}

The spectrum of the 2004 December 27 giant flare from SGR1806-20 was measured
by many different instruments, using many different methods (Hurley et al. 2005,
Boggs et al. 2007, Palmer et al. 2005, Frederiks et al. 2007b).  While they
do not agree on the exact shape of the spectrum, none found evidence for the
existence of a high energy power law component in the Band model.  Our RHESSI-only spectral fits of GRB 051103 
are consistent with this, but in two of the joint fits this component is preferred (table 1).
A blackbody can be fit to the RHESSI data, but only over the 800 keV - 5 MeV range;
kT is approximately 550 keV for this fit, and the chi-square is about 1.5 per degree of
freedom.

Table 2 contains a comparison of the energetics of the giant flares observed to
date.  Because of the uncertainties in the SGR distances, as well as the different energy ranges, time resolutions, and
characteristics of the various instruments which observed these events,
comparisons between the values given are probably uncertain by at least a factor of 3.
The total energy of GRB 051103 at the distance of M81, 7.5$\times$10$^{46}$ erg, does not
strain the possibilities of the magnetar model.  However, PGC 2719634, whose distance is
unknown, also remains a possible host (Lipunov et al. 2005b).

\begin{table*}
\begin{minipage}{140mm}
\caption{
Approximate energies and peak luminosities of the SGR giant flares, and of GRB 051103.  
}
\begin{tabular}{@{}lccc@{}}
\hline
SGR   & Energy, erg &  Peak Luminosity, erg cm$^{-2}$ s$^{-1}$ & Assumed distance, kpc \\
\hline

0525-66\footnote{Mazets et al. 1979} & $\rm 1.2 \times 10^{44} $ & $\rm 5 \times 10^{44}$ & 55 \\
1900+14\footnote{Hurley et al. 1999; Tanaka et al. 2007} & $\rm 4.3 \times 10^{44} $ & $\rm 2 \times 10^{46}$ & 15 \\
1806-20\footnote{Hurley et al. 2005; Terasawa et al. 2005; Frederiks et al. 2007b} & $\rm 2-5 \times 10^{46} $ & $\rm 2-5 \times 10^{47} $ & 15 \\
GRB 070201\footnote{Mazets et al. 2008} & $\rm 1.5 \times 10^{45} $ & $\rm 1.2 \times 10^{47} $ & 780 \\
GRB 051103 & $\rm 7.5 \times 10^{46} $ & $\rm 4.7 \times 10^{48} $ & 3600 \\
\hline
\end{tabular}
\end{minipage}
\end{table*}


\begin{figure}
\includegraphics[scale=0.45]{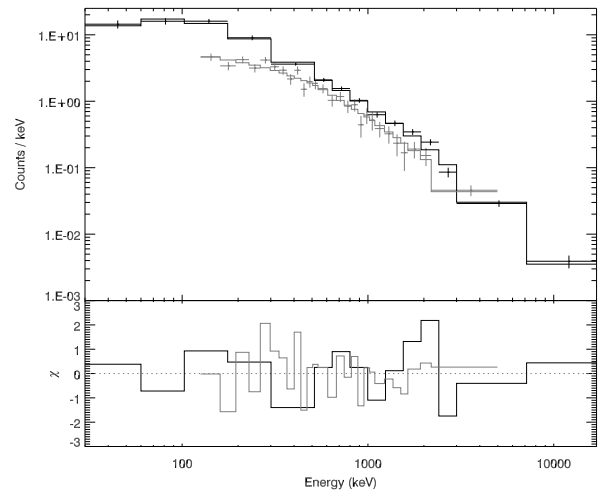}
\caption{Joint spectral fit of RHESSI (black) and Suzaku-WAM (gray) data for
a one-second interval containing the burst.}
 \label{fig-KR23}
 \end{figure}


\begin{figure}
\includegraphics[scale=0.5]{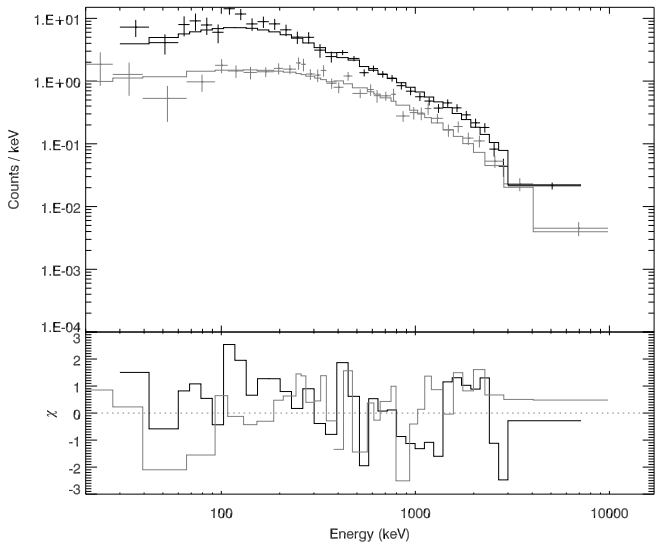}
\caption{Joint spectral fit of RHESSI (black) and Konus-Wind (gray) data for
interval 1.  The overplot model is the best-fit Band function; the 
normalization between the datasets was allowed to vary in the fit.}
 \label{fig-KR1}
 \end{figure}


\begin{figure}
\includegraphics[scale=0.5]{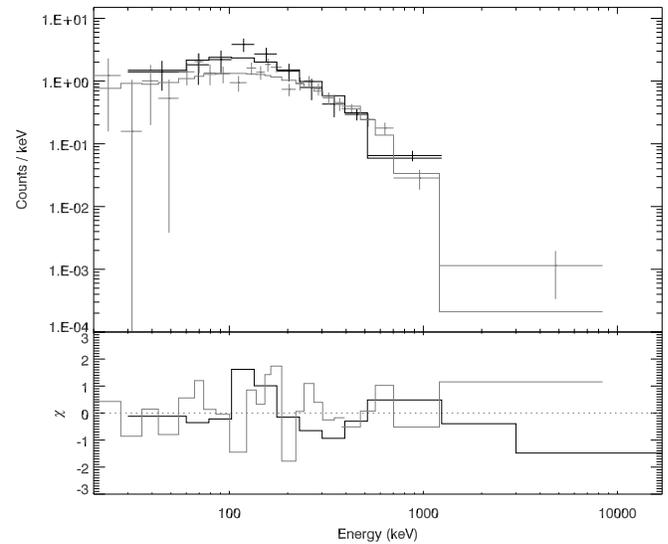}
\caption{Joint spectral fit of RHESSI and Konus-Wind data for
interval 2-3; symbols as in Figure \ref{fig-KR1}.  Two high-energy 
RHESSI points were negative and are omitted from the logarithmic counts plot.}
 \label{fig-KR23}
 \end{figure}

\section{Optical Observations and Analysis}
Observations were obtained on 6$^{th}$ November 2005 using the Mosaic wide field optical 
imaging camera at the KPNO 4m telescope. These data reach a limiting magnitude of 
$\sim24.5$ in the R band, which is considerably deeper than the study completed by \cite{ofek2006}. 
The observations covered the majority of the original error region, 
and in particular that part nearest to the galaxy M81. The images were flat-fielded 
and sky subtracted using standard tasks within {\sc Iraf} \footnote{{\sc Iraf} is distributed by the National
Optical Astronomy Observatories, which are operated by the Association
of Universities for Research in Astronomy, Inc., under cooperative
agreement with the National Science Foundation.}.

For comparison, pipeline-reduced images of the region from CFHTLS were obtained via the 
Virtual Observatory \citep{walton2006}. These formed part of the wide synoptic 
survey in the R band, also to a limiting magnitude of $\sim$25 \citep{ilbert2006}. Coincidentally, 
the region was imaged on 1$^{st}$ November 2005, 2 days prior to the burst, and re-imaged within 
1 month after the burst. This provided an ideal data set for comparison to the KPNO images 
as the timescale between the first images by CFHT and our images from KPNO is only 6 days, 
minimising any modulation in long-period variable stars in the disk/halo of M81.

Figure 1 shows the previous error quadrilateral, 
the refined 3$\sigma$ error ellipse and the fields covered by our KPNO observations, in 
relation to M81. Our observations were positioned to cover the original error quadrilateral 
 but still cover 62\% of the refined 3$\sigma$ ellipse and contain 76\% of the total
likelihood. It is important to note that our observations cover the region closest to M81, 
and therefore our search addresses the possible association of GRB 051103 with M81.

Initially, we searched the images for variability of afterglow counterparts, either at 
the distance of M81 or in the background, by visual inspection and no obvious afterglow 
candidate was found. The magnitudes of sources within these images were then studied 
using {\sc SExtractor} within {\sc Gaia}. They were all calibrated to the r band 
magnitudes of stars in the surrounding region as published in the Sloan Digital Sky Survey 
(SDSS) \citep{adelman2008}. The r band filter used by CFHT matched the filter used in SDSS 
and the filter used by KPNO was a Cousins R band filter. Although this is partially taken 
into account in the calibration to r band magnitudes, there are some sources which have large colour 
differences, for example very red sources. If a source appeared to differ 
in magnitude between the CFHT and KPNO images, the colour correction was calculated using a 
formula developed by \cite{lupton2005} and it was then determined if the magnitude difference 
was due to colour effects. If it was not due to colour effects, the source was investigated 
further. It is important to note that there may be a source within the field which was varying 
but has not been identified due to this colour correction method. However, this method would only 
miss objects with a variability of $\leq$0.3 magnitude (the average colour correction factor used).

Although the magnitude of some stars differed between the images, these were found to be caused 
by other factors, for example, being near chip edges or large diffuse galaxies unidentified 
by {\sc SExtractor}. One of the stars in the region studied has a varying magnitude 
on the images studied and further investigation confirmed it is likely a variable star.

We checked extended sources to look for a conventional SGRB afterglow within a moderately 
distant host galaxy, with a limiting magnitude of $\sim23.3$. If an extended source appeared 
to be varying due to a possible point source being superimposed on it, the colour correction 
was calculated and the object was studied in more depth by eye. This involved using the software 
to match seeing conditions and measure the size of the object, and then to check if there was an 
indication of a change in shape which might indicate a superimposed afterglow component.

In addition to the photometry described above we also searched for
afterglow candidates with PSF-matched image subtraction, using a
modified version of the ISIS code \citep{alard1998, alard2000}.
This method gives us a better chance of finding sources that are blended
with other, brighter objects (i.e. bright host galaxies).
After cosmic ray cleaning and resampling on a common pixel grid, we
subtract the KPNO data with the CFHT data taken before and with the data
taken after the burst. We found no credible afterglow candidates.

The analysis of the images found no optical afterglow candidate in the region studied 3 days 
after GRB 051103. This can place constraints on the progenitor of GRB 051103 by considering 
the expected results for the potential progenitors.

\subsection{Progenitor option 1: a Short GRB (SGRB)}

\begin{table}
\begin{center}
\caption{The observed fluence, in the energy band 15-150keV, of SGRBs with observed 
R band magnitudes at approximately 3 days.}
\begin{tabular}{| c | c | c |}
\hline
SGRB & Fluence & R band Magnitude at 3 days \\
   & $10^{-7}\;$erg~cm$^{-2}$ &           \\
\hline
051221A & 11.6$\pm$0.4 $^{(1)}$ & 24.12$\pm$0.28 $^{(2)}$ \\
051227 & 2.3$\pm$0.3 $^{(3)}$ & 25.49$\pm$0.09 $^{(4)}$ \\
060121 & 26.7$^{+5.3}_{-20.2}$  $^{(5)}$ & 25$\pm$0.25 $^{(6)}$ \\
060614 & 217$\pm$4 $^{(7)}$ & 22.74$\pm$0.31 $^{(8)}$ \\
061006 & 14.3$\pm$1.4 $^{(9)}$ & $>$23.96$\pm$0.12 $^{(10)}$\\
070707 & 0.334$^{+0.753}_{-0.316}$ $^{(11)}$ & 26.62$\pm$0.18 $^{(12)}$ \\
070714B & 7.2$\pm$0.9 $^{(13)}$ & $<$25.5 $^{(14)}$\\
071227 & 2.2$\pm$0.3 $^{(15)}$ & $>$24.9 $^{(16)}$ \\
080503 & 20.0$\pm$1 $^{(16)}$ & 25.90$\pm$0.23 $^{(17)}$\\
\hline
\end{tabular}
\end{center}
$^{(1)}$ \cite{cummings2005}
$^{(2)}$ \cite{soderberg2006}
$^{(3})$ \cite{hullinger2005}
$^{(4)}$ \cite{davanzo2009}
$^{(5)}$ an approximate fluence calculated using spectral parameters published by \cite{golenetskii2006}
$^{(6)}$ based on observations by \cite{levan2006a}
$^{(7)}$ \cite{barthelmy2006}
$^{(8)}$ \cite{mangano2007}
$^{(9)}$ \cite{krimm2006}
$^{(10)}$ an upper limit based on observations 2 days after the burst completed by \cite{davanzo2009}
$^{(11)}$ an approximate fluence calculated using spectral parameters published by \cite{golenetskii2007}
$^{(12)}$ \cite{piranomonte2008}
$^{(13)}$ \cite{barbier2007}
$^{(14)}$ a lower limit based on observations 4 days after the burst completed by \cite{perley2008}
$^{(15)}$ \cite{sato2007}
$^{(16)}$ a 3$\sigma$ upper limit published by \cite{davanzo2009}
$^{(17)}$ \cite{ukwatta2008}
$^{(18)}$ \cite{perley2008}
\end{table}

The optical afterglows of various SGRBs have been studied and these data can 
be used to predict the range of afterglow properties of an SGRB of a particular gamma-ray fluence. There 
is evidence for a reasonable correlation, to first order, between gamma-ray fluence 
and afterglow flux \citep{nysewander2008, gehrels2008}. Using {\sc Xspec}, we created a 
model spectrum of GRB 051103, using the RHESSI + KW joint fits in Table 1, 
and estimated the fluence of GRB 051103 in the energy band 15-150 keV to be approximately 
$9.6^{+14.5}_{-3.7} \times 10^{-7}$erg~cm$^{-2}$.

It is possible to compare GRB 051103 to other SGRBs in the BAT catalogue \citep{sakamoto2008} 
using the approximate fluence, calculated for the energy band 15-150 keV, and the photon indices 
given in Table 1. GRB 051103 is isolated at the extreme bright, hard 
end of the SGRBs in the {\it Swift} distribution  \citep[c.f. Figure 14 from][]{sakamoto2008}. 
Similarly, in the study of short bursts by Mazets et al. (2004) over the much wider Konus
energy range, of the 109 spectra which could be characterized
by an E$_{peak}$, none exceeded 2.53 MeV.  The peak energy of GRB 051103 is approximately
3 MeV (table 1).  Thus if GRB 051103 is an SGRB rather than an SGR giant flare, it is a fairly extreme case.

We compared the fluence of this burst to other SGRBs observed by the {\it Swift} Satellite. Table 
3 provides the data of SGRB with fluences in the band 15-150keV and late optical observations, 
obtained from the relevant GCNs, and measured optical afterglows in the R band, approximately 
3 days after each burst \footnote[1]{It is important to note the classification of some of 
these SGRBs are currently being debated \citep{zhang2009}}. For two of the bursts it was necessary to 
estimate the fluence in the correct energy band using the same method as with GRB 051103. 
This is not a complete sample of SGRBs, as there are a number with a relatively low gamma-ray 
fluence that were either not observed optically, were not observed for longer than a few hours, or 
did not have a detected optical afterglow. We chose this sample so we did not have to rely on 
the assumption that we can extrapolate the light curve to later epochs and because they are of 
a similar gamma-ray fluence to GRB 051103. We compared the SGRBs in Table 3 
to GRB 051103 and predict the optical afterglow would have an R band magnitude of $\sim24$ 
as it is at the higher end of the fluence distribution. This is within the limiting magnitude 
of the KPNO and CFHTLS images used, but would have been unobservable in the images obtained 
by \cite{ofek2006}. As no afterglow was observed, this rules out most typical SGRBs in the 
region of the error ellipse covered by our imaging. However, there are cases of SGRBs with extremely 
faint optical afterglows, for example GRB 080503, which had a similar fluence to GRB 051103 
and an r band magnitude of 25.90$\pm$0.23 at 3 days \citep{perley2008}. So the observations 
cannot rule out an unusually faint SGRB in this region similar to GRB 080503. Additionally, GRB 051103 
could be a classical SGRB in the part of the error ellipse not studied in this paper.

\subsection{Progenitor option 2: an SGR giant flare in M81}

Conversely, GRB 051103 could be an SGR giant flare in M81 with similar energy to the giant flare from SGR 1806-20 
\citep{golenetskii2005} and a very faint optical afterglow \citep{eichler2002, levan2008}. 
Using observations of the giant flare from SGR 1806-20, we can predict the apparent optical magnitude of an SGR 
in M81. The distance to SGR 1806-20 has proven difficult to determine; the distance modulus 
adopted by many authors is 15.8 \citep{corbel1997}, although \cite{bibby2008} recently obtained 
a revised distance modulus estimate of 14.7$\pm$0.35mag. Here we continue to use the larger 
distance modulus as this will provide an approximate upper limit on the absolute magnitude. The 
giant flare from SGR 1806-20 had an observed radio afterglow and this has been used by \cite{wang2005} 
to make predictions of the apparent R band magnitude of the afterglow. Their analysis suggests that 
the giant flare would have had an apparent magnitude of $\sim22$ at 3 days, and hence an absolute magnitude of
$ M \approx 6 $. Taking this as the absolute magnitude of any afterglow of GRB 051103 if it 
is an SGR giant flare, and using the distance modulus to M81  
of 27.8 \citep{freedman1994}, we conclude the afterglow would be expected to have an apparent 
magnitude of $> 34$. Despite the many uncertainties involved in this calculation, we can have
some confidence that such an afterglow would not be detectable with the data available. For 
future reference, it is important to note that with more accurate positions and rapid follow up 
observations it may be possible to observe the optical afterglows of extragalactic giant flares. For example, 
if there were a second potential giant flare in M81 we predict the optical afterglow would have a peak apparent K band 
magnitude of $\sim$20 at 86s after the giant flare and would fall to $\sim$26 at 1 hour. This is observable with 
current and upcoming facilities, for example the European Extremely Large Telescope (E-ELT).
However, these predictions are based upon the theoretical models of SGR giant flares being similar to the 
blast wave model used to describe classical GRBs. \cite{wang2005} use the blast wave model and radio 
observations of the giant flare from SGR 1806-20 to extrapolate the optical afterglow. SGRs have been observed 
during periods of activity using ROTSE-I \citep{akerlof2000} and {\it Swift} \citep[for example ][]{atel2127}, 
and IR observations have been obtained for SGR 1900+14 4.1 days after outburst detecting no variability 
\citep{atel26}. These have provided upper limits on the optical afterglows from the softer spectrum, shorter, and weaker bursts seen 
during active phases of SGRs but it is important to note that there have been no reported rapid optical 
follow up observations of galactic SGR giant flares, which have a significantly higher fluence and are spectrally harder 
than these bursts. Therefore, we are completely reliant on theoretical predictions 
and future observations may show discrepancies with these predictions. Indeed, our observations with a 
limiting magnitude of 24.5, giving an absolute magnitude -3.3 assuming it is at a distance of 3.6 Mpc, 
constitute one of the deepest absolute magnitude searches for an afterglow from a possible SGR giant flare. This 
absolute magnitude is only exceeded by the search for an afterglow from GRB 070201, which is a candidate 
SGR giant flare in M31, corresponding to an absolute magnitude of -7.4 obtained 10.6 hours after the burst \citep{ofek2008}. 
However, as we discuss later, it is unlikely that both of these events were SGR giant flares \citep{chapman2008}.
 
From the {\it GALEX} UV imaging \citep{martin2005}, there is evidence that the error 
ellipse does contain star forming regions in the outer disk of M81. The two brightest UV 
sources are marked on Figure 6 \citep{ofek2006}. These young stellar regions in M81 could 
host an SGR which could emit a giant flare. Similarly, these UV regions could be the locations of 
massive star clusters, and SGRs 1900+14 and 1806-20 have been associated with massive star 
clusters \citep{mirabel1999, vrba2000}. However, if GRB 051103 is an SGR giant flare in M81, we might 
also expect to find a young (up to $\sim10^{4}$ years old, \cite{duncan1992}) SNR in the 
nearby region, although this association is still being debated \citep{gaensler2001, 
gaensler2005}. When an SGR is formed, it is theoretically possible that it is given a kick 
of up to $1000\;$km~s$^{-1}$ or more \citep{duncan1992} and therefore could have traveled 
a distance of $>$10pc from the SNR. However, this is only equivalent to an angular separation of 
$\sim0.6$\,arcsec at a distance of 3.6Mpc \citep{freedman1994}. Hence, an accompanying SNR 
would still be expected to fall within the error ellipse. Of the known SNR in M81 
\citep{matonick1997}, there are none within the error ellipse.

M81 has been studied by the {\it Chandra X-Ray Observatory} \citep{swartz2003} and three X-ray 
sources are within the error ellipse. However, they have not been 
identified in visible or radio observations. Additionally, they have 
not been identified with known SNR, nearby stars, are not co-incident with HII starforming 
regions (the expected location of SGRs -- \cite{duncan1992}) and are more likely to be X-ray 
binary systems than unidentified SNR \citep{swartz2003}. This survey had a limiting 
luminosity of $3\times10^{36}$ erg s$^{-1}$, which means it would detect very young supernovae, 
as the oldest supernovae with detected X-ray afterglows had a luminosity of $\sim10^{37}$ 
erg s$^{-1}$ and an age of $\sim60$ years \citep{soria2008}. Additionally, this survey would 
detect the X-ray luminous SNR as these have a luminosity of up to $\sim10^{37}$ erg s$^{-1}$ 
but would not detect the X-ray faint SNRs which have a luminosity of $\sim10^{34}$ erg s$^{-1}$ 
\citep{immler2005}. SGRs are well known to be quiescent soft X-ray emitters and \cite{mereghetti2000} 
have measured the soft X-ray flux of SGR 1806-20 to be approximately $10^{-11}$erg~cm$^{-2}$~s$^{-1}$. 
\cite{frederiks2007} determined that the {\it Chandra Observatory} would be unable to observe 
directly the persistent X-ray flux from an SGR like SGR 1806-20 in M81.

An alternative method of searching for SNR is to use H$\alpha$ and OIII narrow band observations. 
The Isaac Newton Telescope (INT) has been used to search for planetary nebulae in M81 by 
\cite{magrini2001} and they have found 171 potential candidates, some of which are in the nearby 
region of the refined error ellipse. Their criteria for differentiating between an SNR and a 
planetary nebula is that planetary nebulae cannot be spatially resolved and SNR are. A young SNR, as 
required for an SGR, could be misidentified as a planetary nebula by this criterion, since a one 
arcsecond region corresponds to a physical size of $\sim 20$pc. Young SNRs may well
be significantly smaller than this, since an expansion velocity of a few thousand km s$^{-1}$
over a magnetar lifetime of $\sim 10^4$ years leads to sizes of $10-50$ pc. Indeed, 
many SNRs in M82 appear (based on radio maps) to be fairly compact \citep{fenech2008}. However, the 
nearest is still $\sim$23 arcsec from the error ellipse, and we showed earlier that an SGR in M81 
would only be able to travel $\sim$0.6 arcsec from its birthplace. The H$\alpha$ luminosity 
of SNRs in nearby disk galaxies tends to be greater than $\sim10^{36}$ erg s$^{-1}$ \citep{grijs2000} 
and the work by \cite{magrini2001} had a limiting H$\alpha$ flux of less than $6\times10^{-17}$ erg 
cm$^{-2}$ s$^{-1}$ which corresponds to a limiting luminosity of $\sim10^{35}$ erg s$^{-1}$. Therefore, 
we would expect their survey to find a candidate SNR. We used the recently published 
H$\alpha$ and OIII images from the INT Wide Field Camera Imaging Survey \citep{mcmahon2001}, with 
a limiting luminosity of $\sim 10^{35}$ erg s$^{-1}$ at the distance of M81 as these are the same 
images as used by \cite{magrini2001}, and compared them with 21cm radio images from THINGS 
\citep{walter2008} and Chandra X-ray source positions \citep{swartz2003} to search for previously 
unidentified SNRs within the error ellipse. There is a possible circular 21cm radio source coincident 
with a Chandra X-ray source of approximately the right flux for an SNR in M81 \citep[source 15 in][]{swartz2003}. 
However, the 21cm radio source is too large for an SNR of the required age and there is no convincing 
supporting evidence of a source within the other images studied. Using the published known X-ray sources 
we might have expected to find an SNR if it was very young or bright and we would have expected to find an 
associated SNR using the H$\alpha$ images. We identified no convincing associated SNR candidates within 
the error ellipse.

\begin{figure}
\centering
\includegraphics[width=8.2cm]{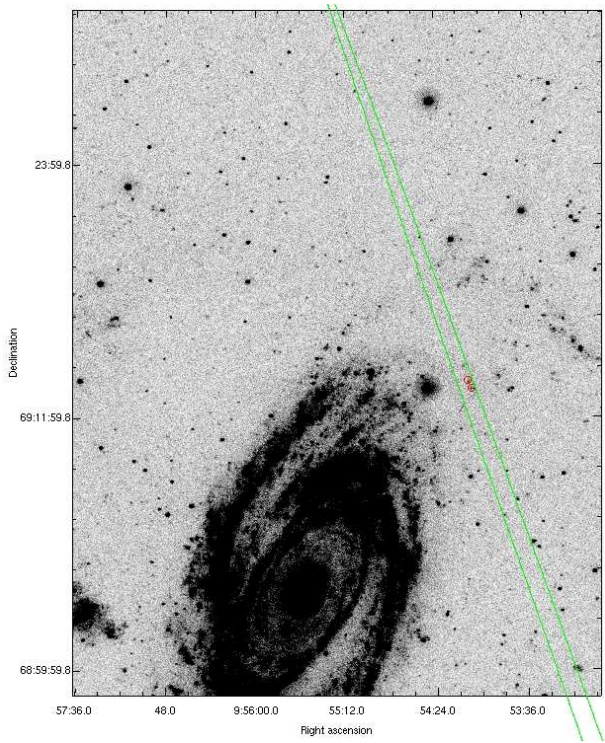}
\caption{This shows an image taken using {\it GALEX} showing the UV sources within the error 
ellipse. Two of the brightest sources discussed by \citet{ofek2006}, within the ellipse, 
are highlighted by the red circles.}
\end{figure}

Although it has been determined that the error ellipse does cross potential star formation 
regions as required by the majority of SGR models, it should also be noted that this is not 
essential for all. An alternative route has been proposed for producing a magnetar by white 
dwarf (WD) mergers \citep{king2001, levan2006b}. As WD have long lifetimes, WD-WD mergers 
would be associated with older populations of stars. It is possible that accretion induced 
collapse (AIC) will drive off a fraction of the envelope, leaving something akin to an SNR 
behind \citep[e.g.][]{baron1987}. The mechanisms underlying AIC are poorly understood, and 
the physical characteristics and detectability of such remnants are not clear. Therefore a 
SGR produced through this channel could be formed in an old stellar population within the 
outer disk or halo, and the non-detection of an SNR within the region does not place 
constraints on this model.

If the progenitor was an SGR giant flare, then there might be significant similarities in the light 
curve and spectrum of GRB 051103 to the giant flare from SGR 1806-20. \cite{ofek2006} noted that the 
light curve of these two events were consistent, i.e. the light curve of GRB 051103 is 
similar to what would be expected from an extragalactic version of the giant flare from SGR 
1806-20. In Table 1, we have shown, for the joint RHESSI + KW fits, that initially $\alpha = 0.13^{+0.14}_{-0.11}$ 
and softens to $\alpha = 0.39^{+0.35}_{-0.30}$. Although this is unusually hard for a GRB, it is consistent 
with the photon index of the giant flare from SGR 1806-20,  $\alpha = 0.2\pm0.3$ 
\citep{palmer2005}. The peak luminosity of GRB 051103, assuming it was from an SGR in M81, is 
approximately $4.7 \times 10^{48}$ erg s $^{-1}$. This is a factor of 10 brighter 
than the peak luminosity of the giant flare from SGR 1806-20, which is $2-5 \times 10^{47}$ erg s$^{-1}$ 
assuming it is at a distance of 15kpc \citep{hurley2005}. With the revised distance estimate from 
\cite{bibby2008}, the peak luminosity of the giant flare from SGR 1806-20 would be $7 \times 10^{46}$ erg s 
$^{-1}$, suggesting that a much smaller percentage of SGRBs are SGR giant flares. This value is 30 times 
fainter than the peak luminosity of GRB 051103 if it was from an SGR giant flare in M81 and in this case 
GRB 051103 would be the most luminous SGR giant flare observed. In comparison, the peak luminosity of GRB 
070201 is $1.14 \times 10^{47}$ erg s $^{-1}$ assuming it was in M31 \citep{ofek2008}, which is 
an order of magnitude fainter than GRB 051103 and comparable to the giant flare from SGR 1806-20. It is 
important to note however, that there is currently no theoretical upper limit for the energy 
of a giant flare. \cite{duncan1992} showed that the total energy available is given by 
$E~\propto~3\times10^{47}~B_{15}^{2}~erg$ where $B_{15}=B/10^{15}~G$. Therefore, the magnetic 
dipole (B) of SGR 1806-20 would only need to increase by a factor of $\sim$5 to produce a giant flare 
with an energy that is 30 times greater than the one from SGR 1806-20.

Although the gamma-ray data suggest that GRB 051103 may be an extragalactic SGR giant flare, it is 
important to note that SGR giant flares are rare events. Considering plausible luminosity functions,
\cite{chapman2008} calculated the probability 
that the IPN would observe a giant flare, with energy greater than the energy emitted by the giant flare from 
SGR 1806-20, in the region surveyed during the 17 years it has operated. For one giant flare, they 
calculated the probability to be 10\%. However, as we discussed in the introduction, there are 
four potential candidates for extragalactic SGR giant flares, including GRB 070201 near M31 which has 
been identified as an SGR giant flare by \cite{mazets2008}. The probablility that the IPN has detected 
two SGR giant flares, with energy greater than the giant flare from SGR 1806-20, is 0.6\% 
\citep{chapman2008}. Recently, several new SGR candidates have been identified including 
0501+4516, 1550-5418 and possibly 0623-0006 \citep{barthelmy2008a, krimm2008, barthelmy2008b}, 
which may imply that the number of SGRs in the Milky Way is higher than previously thought. 
In this case, the luminosity of the giant flare from SGR 1806-20 would have to be at the peak of the 
luminosity function of SGR giant flares and therefore giant flares of this luminosity must be extremely rare events.
This argues that GRB 051103 is unlikely to be a second SGR giant flare in the nearby universe.

\section{Conclusions}

GRB 051103 illustrates the difficulties of identifying a short burst
as an extragalactic giant magnetar flare beyond a reasonable doubt.  Even setting aside the questions of detecting and
localizing such events, and establishing their associations with nearby galaxies, their interpretation is problematic.  On the one hand, the localization,
short duration, and hard energy spectra of GRB 051103 suggest
that it is a giant flare from M81.  However, a deeper analysis of its time history demonstrates that
the periodic component, which is a key signature of giant flares, is unlikely to ever be detected at
great distances by the IPN if all giant flares are similar to the three observed to date.  The energy spectrum at the peak of the emission is very hard (E$_{peak}\sim$3 MeV), and is
detected to 7 MeV
at the 3$\sigma$ level, with marginal emission up to 17 MeV.  Yet it is not inconceivable that a short duration
GRB could have these properties.  Although the E$_{peak}$ of GRB 051103 evolves from hard to soft, the evolution to a very soft spectrum, which is expected during
the oscillatory phase of an SGR giant flare, is undetectable, as is the oscillatory phase itself.  Thus evidence for an extragalactic giant flare origin
of GRB 051103 remains tantalizing, but inconclusive.   On a more positive note, if an extragalactic magnetar flare occured within
the Swift-BAT field of view, so that the XRT could begin observing within a minute or so, the periodic component would
be detectable at low energies to at least 10 Mpc (Hurley et al. 2005). 

We have presented new optical observations of GRB 051103 and have determined that there is no R band optical 
afterglow with a limiting magnitude of $\sim24.5$ (for an afterglow overlapping a host galaxy, 
the limiting magnitude is $\sim23.3$) in the region of the error ellipse covered by our 
observations.  Comparison of the prompt emission of GRB 051103 with a sample of other SGRBs leads us to
conclude that if it was a classical SGRB we would expect to have located an optical afterglow in our 
observations.

In contrast, if GRB 051103 were an SGR giant flare in M81, non-detection of an afterglow would not be 
surprising as the expectations for optical afterglow emission lie significantly below the 
limits obtained here, or the limits likely to be attained via current technology. The case for an SGR 
origin would be strengthened if there were an accompanying SNR within the error ellipse, but there 
is no evidence of this. An SGR produced via accretion induced collapse of a WD \citep{levan2006b} 
would, however, remove the requirement for an SNR. Additionally, the luminosity of GRB 
051103, assuming it is from an SGR giant flare in M81, is significantly higher than known SGR giant flares but 
still attainable with current theoretical models. Giant flares with luminosity similar to the giant flare from 
SGR 1806-20 are extremely rare and it is unlikely that GRB 051103 and GRB 070201 are both 
extragalactic SGR giant flares.

Although we have not considered this option in detail, it is possible that the progenitor 
of GRB 051103 was a compact binary merger in M81. In this case, it would just be within the reach of 
current gravitational wave searches. This scenario was ruled out at $>$99\% confidence for GRB 
070201 in M31 using the Laser Interferometer Gravitational-wave Observatory (LIGO) observations, 
and distances out to 3.5 Mpc were ruled out to 90\% confidence \citep{abbott2008}. The LIGO Scientific 
Collaboration is currently considering a search for gravitational-wave signals in the data surrounding 
GRB 051103 (G. Jones and P. Sutton, private communication).

\section*{Acknowledgments}

KH is grateful for IPN support from the NASA Guest Investigator programs for
Swift (NASA NNG04GQ84G), INTEGRAL (NAG5-12706), and Suzaku (NNX06AI36G); 
for support under the Mars Odyssey Participating Scientist Program, JPL
Contract 1282046; and under the HETE-II co-investigator program, MIT Contract
SC-A-293291.  We are also grateful to the Konus-Wind team - E. Mazets,
S. Golenetskii, D. Frederiks, V. Pal'shin, and R. Aptekar -
for contributing the Konus data to this study.

AR, KW and ER, NRT \& AJL would like to acknowledge funding from the Science and Technology 
Funding Council.

This research has made use of data obtained using, or software provided by, the UK's 
AstroGrid Virtual Observatory Project, which is funded by the Science and Technology 
Facilities Council and through the EU's Framework 6 programme.

The National Optical Astronomy Observatory (NOAO) consists of Kitt Peak National
Observatory near Tucson, Arizona, Cerro Tololo Inter-American Observatory near La 
Serena, Chilie, and the NOAO Gemini Science Centre. NOAO is operated by the 
Association of the Universities for Research in Astronomy (AURA) under a cooperative 
agreement with the National Science Foundation.

Based on observations with MegaPrime/MegaCam, a joint project of CFHT and CEA/DAPNIA, 
at the Canada-France-Hawaii Telescope (CFHT) which is operated by the National Research 
Council (NRC) of Canada, the Institut National des Sceicne de l'Univers of the Centre 
National de la Recherche Scientifique (CNRS) of France, and the University of 
Hawaii. This work is based in part on data products produced at TERAPIX and the 
Canadian Astronomy Data Centre as part of the Canada-France-Hawaii Telescope Legacy 
Survey, a collaborative project of NRC and CNRS.

The Digitized Sky Surveys were produced at the Space Telescope Science Institute under 
U.S. Government grant NAG W-2166. The images of these surveys are based on photographic 
data obtained using the Oschin Schmidt Telescope on Palomar Mountain and the UK Schmidt 
Telescope. The plates were processed into the present compressed digital form with the 
permission of these institutions.

Based on observations made through the Isaac Newton Group's Wide Field Camera Survey Programme 
with the Isaac Newton Telescope operated on the island of La Palma by the Isaac Newton Group 
in the Spanish Observatorio del Roque de los Muchachos of the Instituto de Astrofísica de Canarias.

This work made use of THINGS, 'The HI Nearby Galaxy Survey' \citep{walter2008}.

We also acknowledge useful discussions with G. Jones and P. Sutton.

\bsp

\label{lastpage}

\end{document}